# The Journal of Space Safety Engineering

## Space debris through the prism of the environmental performance of space systems: the case of Sentinel-3 redesigned mission

--Manuscript Draft--

# Space debris through the prism of the environmental performance of space systems: the case of Sentinel-3 redesigned mission

Thibaut Maury [(1)], Sara Morales Serrano [(2)], Philippe Loubet [(1)], Guido Sonnemann [(1)], Camilla Colombo [(3)], and Luisa Innocenti [(4)]

[(1)] Univ. Bordeaux, CNRS, Bordeaux INP, ISM, UMR 5255, F-33400 Talence (France)
[(2)] European Space Agency, ESTEC, Keplerlaan 1, NL-2201 AZ Noordwijk (The Netherlands)
[(3)] Politecnico di Milano, Department of Aerospace Science and Technology, I-20156 Milan (Italy)
[(4)] European Space Agency, HQ, Rue Mario Nikis 8-10, F-75015 Paris (France)

**ABSTRACT**

Like any industry, space activities generate pressures on the environment and strives towards more sustainable activities. A consensus among the European industrial stakeholders and national agencies in the Space sector is emerging on the need to address eco-design through the prism of the environmental Life Cycle Assessment (LCA) methodology. While the use of LCA is being implemented within the sector, the current scope disregards the potential environmental impact in term of debris generated by space missions on the orbital environment. The paper highlights the relevance of applying LCA holistically during the design phase of space systems, considering potential impacts occurring in the orbital environment during the *utilisation* and *disposal stages* of a space mission. Based on the comparison of two mission designs, the aim is to consider potential emission of space debris into the LCA framework as a way of measuring the resource security for orbits and potential environmental impacts occurring in case of collision.

## 1 INTRODUCTION

### 1.1 Environmental life cycle assessment (LCA) for space missions

As highlighted during the COP21 in Paris, of the 50 essential variables used to assess Earth's climate, 26 are monitored from satellite observations [1]. Though, regarding environmental legislation, the industrial stakeholders of the space sector are not targeted by specific international binding commitments. Nevertheless, environmental performance has become a criterion in purchasing decisions.

Considering guidelines for the evaluation of environmental impacts of space activities, several actors of or related to the European space industry, such as the European Space Agency (ESA) and its associated 'Clean Space initiative' or ArianeGroup, have identified environmental Life Cycle Assessment (LCA), according to ISO14040/44, as the most appropriate methodology to measure and minimise their environmental impact [2]. Since the last three decades, the environmental LCA methodology is considered as a relevant methodology to support decision-makers in the evaluation of the environmental impacts linked to the design, manufacturing, transporting, and disposing of the goods and services [3]. It compiles and evaluates the inputs, outputs and the potential environmental impacts of a product system throughout its life cycle. As a multi-criteria methodology, LCA studies avoid the 'burden-shifting pollution' which consists of transferring impact from an environmental impact category to another, or from a life cycle stage to another. LCA shows how a specific functionality can be achieved in the most environmentally friendly way among a predefined list of alternatives, or in which parts of the life cycle it is particularly important to improve a product to reduce its environmental impacts.

Contractual requirements have been placed in ESA's funded projects to take into account their environmental impacts and promote the development of 'green' technologies through the implementation of the Life Cycle Assessment (LCA) methodology. In particular, an LCA study of the new Ariane 6 launcher in exploitation phase is currently performed [4]. A requirement to perform such assessment from early phases of the project has also been included for two competing Earth Explorer 9 concepts (i.e. FORUM and SKIM) as well as for the standard platform for all future Sentinel missions in the frame of the Copernicus program.

However, space systems deal with a strong particularity, which adds new aspects regarding the scope of the LCA framework (see Figure 1). Rocket launches are the only human activity that crosses all segments of the atmosphere and stays "out" of the natural environment. Environmental impacts of space systems could occur not only in the conventional *Earth environment* but also in the outer space, further referred to as the *orbital environment*.

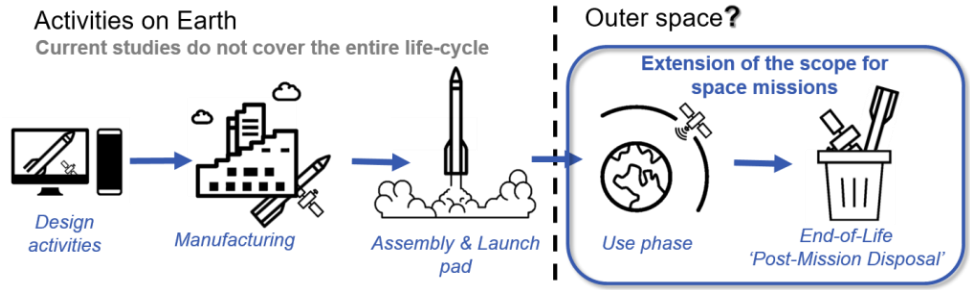

*Figure 1. Life cycle of a space mission. Potential impact regarding debris proliferation in the orbital environment is highlighted.*

Consequently, the scope of environmental LCA for space missions should be extended to cover the phases dealing with in-orbit operations, i.e. the use phase and the end-of-life (EoL) including post-mission disposal of the space missions. By analogy to conventional environmental impacts, the potential generation of fragments in the *orbital environment* is considered as an emission of an environmental stressor damaging the orbital resource. In this way, the potential impacts occurring in the orbital environmental should be assessed to provide a holistic analysis of space missions.

### 1.2 Objectives and outline of the paper

The objective of this paper is to demonstrate the relevance of integrating space-debris related impacts in the methodological framework of environmental LCA for space missions (see Figure 2). For such purpose, the LCA space debris indicator previously developed by Maury et al. [5] in compliance with the LCA standards [6] will be described and applied to a theoretical case study. For the first time [2], we propose to merge this approach dealing with space debris related impacts and the conventional Earth environmental impacts computed in LCA (such as climate change, toxicity, *etc*.) for a space mission.

The chosen study case is the space mission of Sentinel-3 from the ESA, as its LCA has already been carried out in the frame of an eco-design study called 'GreenSat'[7–9]. In this study, two design options concerning the choice of propellant were assessed: the use of (i) Hydrazine and then, (ii) LMP-103s as an eco-design measure [10]. This alternative solution would reduce obsolescence risks (due to the European REACH regulation) and could reduce the time the satellite needs to spend in the clean rooms. The LCA showed that the use of alternative LMP-103s could also decrease conventional environmental impacts such as climate change. Furthermore, the choice of propellant, due to its higher theoretical efficiency, also implies the possibility of re-entry of the spacecraft at the end-of-life life (with the same size of propellant tanks) and therefore its associated exposure to space debris. Thus, it is considered as an appropriate study case to include space debris related impacts in the LCA.

After briefly presenting the LCA structure, the goal & scope of the study is defined with a particular focus on the two scenarios described in the 'GreenSat' study. Then, the inventory data collection and the impact assessment methods, including the space debris related impact used in the frame of this study are presented. Based on the previous system definition and calculation steps, the results obtained are exposed. Finally, the main outcomes, challenges and opportunities regarding the environmental impact assessment of space missions are exposed and discussed.

## 2 MATERIAL & METHODS

LCA is carried out in four distinct phases as defined in ISO 14040/44. It starts with an explicit description of *(i)* the goal and scope of the study before providing *(ii) an inventory of flows* from and to the environment for a product system during the life cycle inventory (LCI) phase. Inventory analysis is followed by *(iii) impact assessment* during the life cycle impact assessment phase (LCIA). This phase of LCA aims at evaluating the significance of potential impacts based on LCI flow results. Finally, *(iv)* the *interpretation phase* is based on the identification of the significant issues, limitations, and recommendations and shall be integrated systematically at each step of the LCA study.

### 2.1 Goal & Scope of the LCA study

**Goal of the study.** The goal of the study is to compare the environmental impacts, including space debris related impacts related to two design alternatives of the Sentinel-3B mission.

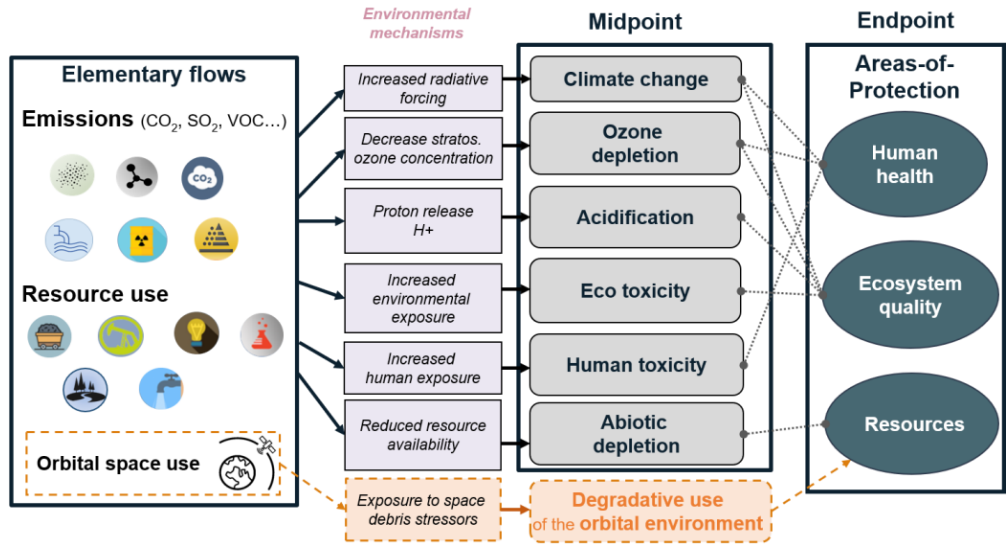

*Figure 2 - Life cycle impact assessment framework. A new impact category is added for LCA of space mission, corresponding to the potential generation of fragments in the orbital environment after a break-up (only collision risk is considered).*

**Scope of the study.** The original Sentinel-3 LCA study was performed following requirements of ESA Space system LCA guidelines [11] which require to consider all segments related to a space mission. Here, we only focus our analysis on *space & ground segments* as presented in Figure 3.

The launcher represents around 99% and the spacecraft 1% of the mass of a complete space mission. Therefore, the environmental impacts related to the launcher (production, launch campaign, launch event) is the major hotspot [12]. However, in a first approach, we focus on the *design, production, utilisation and disposal phases* of the spacecraft only because the ecodesign options are exclusively targeting specific design choices related to the spacecraft. The potential mass increase due to design changes is considered as negligible and do not influence the launcher product system. The infrastructures are also disregarded because the ecodesign practices related to this segment are more in line with facility management optimisation, for instance, site management actions regarding the utilities, than assessment of space systems design and manufacturing processes.

Considering the *utilisation phase,* the nominal operational lifetime for Sentinel-3 is expected to be 7,5 years. However, the amount of embedded propellant reaches 120 kg of hydrazine and can cover an extended timespan, ensuring on-orbit operation for 12,5 years. The expected *Post-Mission Disposal* (PMD) scenario is an uncontrolled re-entry thanks to the decrease of the perigee.

**Description of the scenarios.** In the frame of the GreenSat project, the replacement of Hydrazine propellant embedded in Sentinel-3B was identified as a potential ecodesign improvement. Due to its high human toxicity potential, the hydrazine is currently targeted by the European REACH regulation on chemical substances. It implies strong safety measures during handling and loading of the propellant into the spacecraft as well as the cleaning operations. The authors of the GreenSat study suggested replacing Hydrazine ($N_2H_4$) propellant by LMP-103s (composition: ~ 60% of Ammonium di-Nitramide, ~20% Methanol, ~6% of Ammonia as well as water for mass balance). In term of performance, LMP-103s features a 6% higher specific impulse (Isp) than Hydrazine (235s vs 220s) and a 24% higher density in liquid form (1.24kg/L vs 1kg/L), meaning that more propellant can be stored within the same tank volume. Therefore, we define the following scenarios:

- Baseline scenario: use of Hydrazine as propellant
- GreenSat redesigned scenario: use of LMP-103s as propellant

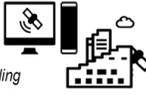

|  | Space Segment | Launch Segment | Ground Segment | Support segment (Infrastructures) |
|---|---|---|---|---|
| Phase A - Feasibility<br>Phase B - Preliminary Definition | Office work and travelling<br>Qualification and testing | *Out of scope* |  | Design facilities |
| Phase C - Detailed Definition<br><br>Phase D - Qualification and Production | Production and development testing of critical elements and engineering models<br>Production of spacecraft components and propellants<br>Qualification, testing and verification<br>Spacecraft assembly | Production of launcher components and propellants<br>Stage assembly |  | *Out of scope*<br>Production & testing facilities |
| Phase E – Utilisation<br>E1 – Launch & Commissioning<br>E2 – Use Phase | Launch campaign | Launch campaign<br>Launch event (launch pad operations and lift-off) | Launch and early operation phase + Commissioning<br>Mission control<br>Energy consumption of payload datacenter | Spaceport<br>Telemetry, Tracking and Command Ground Station(s)<br>Flight Operation Control center(s) |
| Phase F – Disposal | Disposal of the spacecraft | Disposal of the launcher stages | Ground operations for the end of life of the spacecraft |  |

*Figure 3. System boundaries of a space mission according to the specific segments of a space mission based on [13]. For this specific study on Sentinel-3B, only Space & Ground segment were included in the scope. In the space segment, two different propellants are studied (Hydrazine vs LMP). Also, the disposal phase of the spacecraft is further explored in this paper.*

We analyse the delta-V budget needed to performed a 12,5-year mission followed by potential PMD manoeuvres to be considered during the *disposal stage*. The same approach was already proposed in [5,14] for Sentinel-1A spacecraft. We consider equivalent characteristics for both scenarios described above apart from the propellant used. It should be noted that we assume a negligible mass change for the propulsion system as stated in [10]. The potential PMD options are proposed hereafter in Table 1 based on the OSCAR tool simulations of the ESA-DRAMA v3 software [15].

*Table 1. Potential post-mission disposal options to be considered for Sentinel-3 spacecraft*

| Sentinel-3 | Tot. ΔV budget (m/s) | ΔV 12,5-year mission (m/s) | ΔV for direct deorbiting (m/s) | Budget for delayed reentry < 25 years (m/s) | Remaining ΔV for final boost (controlled reentry) |
|---|---|---|---|---|---|
| **Baseline**<br>*120 kg Hydrazine* | 238 | 200 | 203<br>*(too high for both scenarios)* | 77<br>*(too high for the baseline sc.)* | 0 |
| **Alternative**<br>*165 kg LMP-103s* | 342 | | | | 65<br>*(not enough for full controlled reentry)* |

**Functional unit.** The functional unit corresponding to space missions has been debated since the beginning of the LCA application to the space sector. Most of the time, a space mission is designed for a unique purpose (Earth observation, telecommunications, science). For this reason, obtaining a functional unit which enables the comparison based on the 'function' of a satellite is challenging guidelines [11]. To compare the environmental impacts of both space systems including manufacturing, operational lifetime in orbit (i.e. use phase) and PMD phase, the following functional unit was chosen: *"Complete a 12,5-year mission for the Sentinel-3 spacecraft including the deorbit from its operational orbit (inc=98°, h=800 km) to the upper part of the atmosphere (h=120 km)"*.

### 2.2 Life cycle inventory (LCI) data collection

**Conventional LCI.** The 'GreenSat' study is based on a previous ESA-funded LCA study performed by Deloitte [8] for which four iterations in terms of industrial data collection were conducted. A fifth iteration was performed in the frame of 'GreenSat' by Thales Alenia Space (as prime contractor) and Deloitte to complete and improve the inventory model with specific industrial data. Also, the recent development and use of the ESA space-specific LCI database [16,17] brings an added-value, particularly dealing with propellant manufacturing. To answer the functional unit, the following propellant quantities (based on Table 1), are needed: (i) Baseline scenario, use of all the remaining hydrazine corresponding to 17 kg; (ii) GreenSat redesigned scenario, use of around 35kg of LMP-103s to perform a 25-year reentry.

**LCI parameters for 'degradative use of orbits'.** The inventory corresponds to the orbital surface occupied by the product system under study (i.e. the spacecraft) multiplied by its respective on-orbit lifetime, expressed in $m^2 \cdot years$. In addition, the launch mass of the spacecraft shall be considered. The inventory variables which correspond to design parameters are given by the Eq. 1.

$$Inventory = A_c \cdot M \cdot \sum_{Orbits} t_i \qquad [m^2 \cdot kg \cdot yrs] \qquad \text{Eq. 1}$$

$A_c$ is the *average cross-section area of the S/C*. $M$ is the *launch mass* of the spacecraft (in kg). The dwelling time in orbit $t_i$ is mainly dependent on the mission lifetime and the area-to-mass ratio which allows quantifying the effect of orbital mechanics perturbations. Besides, the mass of the spacecraft is also the main parameter involved in the calculation of the number of debris generated when a break-up occurs. $\sum_{Orbits}(t_i)$ expresses the sum of the dwelling time into each orbital cell *i* crossed by the trajectory of the spacecraft. This on-orbit lifetime covers the nominal time of the mission (use stage) plus the post-mission disposal duration representing the End-of-Life (EoL) phase.

The trajectories for the mission lifetime and the theoretical PMD scenarios, as described in Table 1, are propagated thanks to the OSCAR tool of the ESA-DRAMA software v3 [15]. The results are shown in Figure 4.

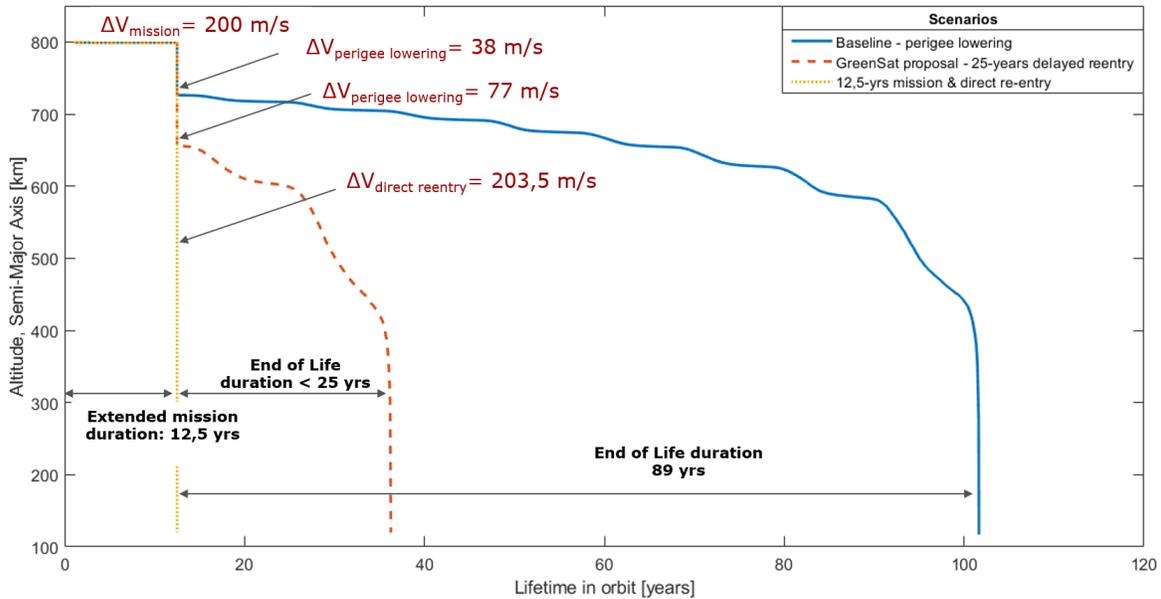

*Figure 4. Semi-major axis of the spacecraft during the operational time of the mission and potential post-mission disposal obtained using the ESA-DRAMA software [15]. The delta-V values previously calculated for the PMD manoeuvres are indicated. The direct re-entry scenario is a discarded option because of the delta-V budget required.*

### 2.3 Conventional life cycle impact assessment (LCIA): selected indicators

The selected LCIA indicators for ESA funded studies are broadly described in the ESA Handbook [11]. Most of them are based on the ILCD 2011 recommended methods [18]. Due to confidentiality aspects, we only present in this paper the relative impact contributions of the processes for the five indicators listed hereafter:

− *Global Warming Potential* (in tonnes $CO_2$ eq.) retrieved from CML2002 impact method [19] and based on the greenhouse gas quantification method proposed by the Intergovernmental Panel on Climate Change [20].
− *Freshwater Aquatic Ecotoxicity Potential* (in disappeared fraction of species) and *Human Toxicity Potential* (in disability-adjusted life years) which quantify the severity of observed effects of substances emitted respectively in an ecosystem integrated over area and time and, on human health, giving more weight to death and irreversible problems than to reversible temporary problems (e.g. a skin or respiratory irritation). Both indicators are based on the USEtox consensus model [21].
− *Fossil Resources Depletion Potential* (in giga-joules of fossil energy consumed) and *Mineral Resources Depletion Potential* (in kg Sb eq.) represent the *Abiotic Depletion Potential* of the CML2002 method [19]. This indicator characterises the depletion of energetic and mineral resources based on the extraction rate and the reserve base estimates.

It should be noted that the *Ozone Depletion Potential* (ODP) is a key indicator in the space sector as the space activities are the only ones responsible for direct emissions within the ozone layer. Previous ESA LCA studies [12] showed that the launch event (which is part of the launch segment) is by far the main contributor on this indicator, i.e. close to 100%, of the impact for a complete space mission scope. Since the launch segment is not included in the scope of this paper, we choose to disregard this impact category.

### 2.4 Specific LCIA indicator for 'degradative use of the orbital resource'

The methodology associated with the development of the LCIA indicator for space debris is widely addressed in [5,22]. We propose here to present the equations leading to the characterisation factors (CFs) presented in Figure 6. These CFs express the potential generation of space debris from a spacecraft, and are discretised for each orbital cell. CFs are computed as the product between (i) the *exposure factor (XF)* that depicts the risk that the spacecraft encounters a space debris, (ii) a $\alpha$ coefficient that depicts the number of debris generated by a collision event depending of the mass of spacecraft and (iii) the *severity factor (SF)* that depicts the fate of the potential fragments generated (Eq. 2).

$$CF_i = XF_i \cdot \alpha \cdot SF_i \quad \text{[potential fragments·years·kg}^{-1}\text{]} \quad \text{Eq. 2}$$

**Exposure factor (XF).** We define the *exposure factor* ($XF_i$) in Eq. 3 as the average flux of space debris passing through a targeted circular orbit $i$ of the LEO region for one year. The orbital *stress* caused by space debris should be assessed for the LEO region to obtain spatially differentiated factors since each orbit presents a different state which allows to classify and differentiate them accordingly. It is done by computing the flux of the catalogued objects in each LEO orbits as done in previous studies [23–25]. It represents the background population (i.e. explosion and collision fragments, rocket bodies, dead and active spacecraft, *etc.*).

$$XF_i = \overline{\phi}_{h,inc,t} \quad [\#.m^{-2}.yr^{-1}] \quad \text{Eq. 3}$$

where $XF$ is the exposure factor for particular circular orbit $i$; $\overline{\phi}$ is the relative flux of catalogued particles provided by the ESA's reference model MASTER-2009 at a given altitude *(h)*, inclination *(inc)* and interval of time *(t)* averaged on a 35-year period based on a 'business-as-usual' perspective. The 35-year period has been chosen with the aim of covering the orbital lifetime of a satellite completing a 10-year mission and a 25-year Post-Mission Disposal as required by the international standard [26]. All debris whose size is higher than 1 cm is accounted for. However, the size of the debris taken into account during the mission lifetime (*i.e. utilisation phase*) is only $1\ cm < XF_{i,mission} < 10\ cm$ representing non-trackable objects for which collision avoidance maneuver is not possible.

**Alpha coefficient (α).** Using the NASA break-up model [27,28], it is possible to express the number of fragments >10 cm generated from a collision per kg of spacecraft depending on the initial *launch mass.* The relation is presented in Figure 5. Since the number of fragments/kg depends on the mass of the spacecraft, we propose to consider a fixed ratio (α coefficient) for a given class of spacecraft (between 1.000 and 1.500 kg as for Sentinel-3): 0.86 fragments/kg.

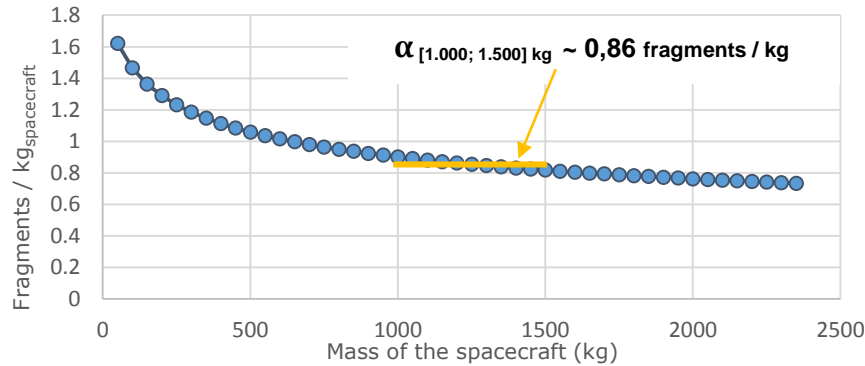

*Figure 5. Fragments > 10 cm released per kg of spacecraft according to the mass of the spacecraft using the NASA break-up model [27,28]. For sentinel-3 the class [1.000; 1.500] is chosen.*

**Severity factor (SF).** According to [29], the percentage of fragments > 10 cm released at an altitude *h (km)* and still on-orbit after a given time *t* (yrs) follows the Eq. 4:

$$P(t,h) = e^{-\frac{t}{128.3 - 0.585892 \cdot h + 0.00067 \cdot h^2}} \qquad [\%] \qquad\qquad Eq.\ 4$$

Where $P$ is the percentage of fragments (in %) still in orbit after a period $t$ (in years) and h is the initial altitude of release (in km).

The cumulative residence time of debris into orbits is obtained by the integral of $P(t,h)$ over a given interval of time. Here, we choose the following time interval: [0:200] yrs. The polynomial part of the Eq. 4 is later expressed as $\tau$ and can be considered as a constant in the integral which is only time-dependent. Thus, the *severity factor* ($SF_i$) for a break-up occurring in given orbit $i$, is given in Eq. 5. It represents the cumulative survivability of a fragment with respect to its altitude of emission expressed in years.

$$SF_{i, 200\ yrs} = \int_{0\ yr}^{200 yrs} e^{-\frac{t}{\tau}} = \left[-\tau \cdot e^{-\frac{t}{\tau}}\right]_{0\ yrs}^{200\ yrs} \qquad [\text{years}] \qquad\qquad Eq.\ 5$$

**CFs and computation of impact score (IS).** Combining Eq. 3, Eq. 5 and the alpha coefficient, we obtain the CFs calculated for a given circular orbit $i$ and presented in Figure 6.

For a specific spacecraft (i.e. product system) with a defined mission and PMD scenario (see Figure 4 & Eq. 1), we obtain the following impact score (*IS*) presented in Eq. 6:

$$IS_{mission} = A_c \cdot M \cdot \sum_i^{orbits} t_i \cdot CF_i \qquad [\text{potential fragments·years}] \qquad\qquad Eq.\ 6$$

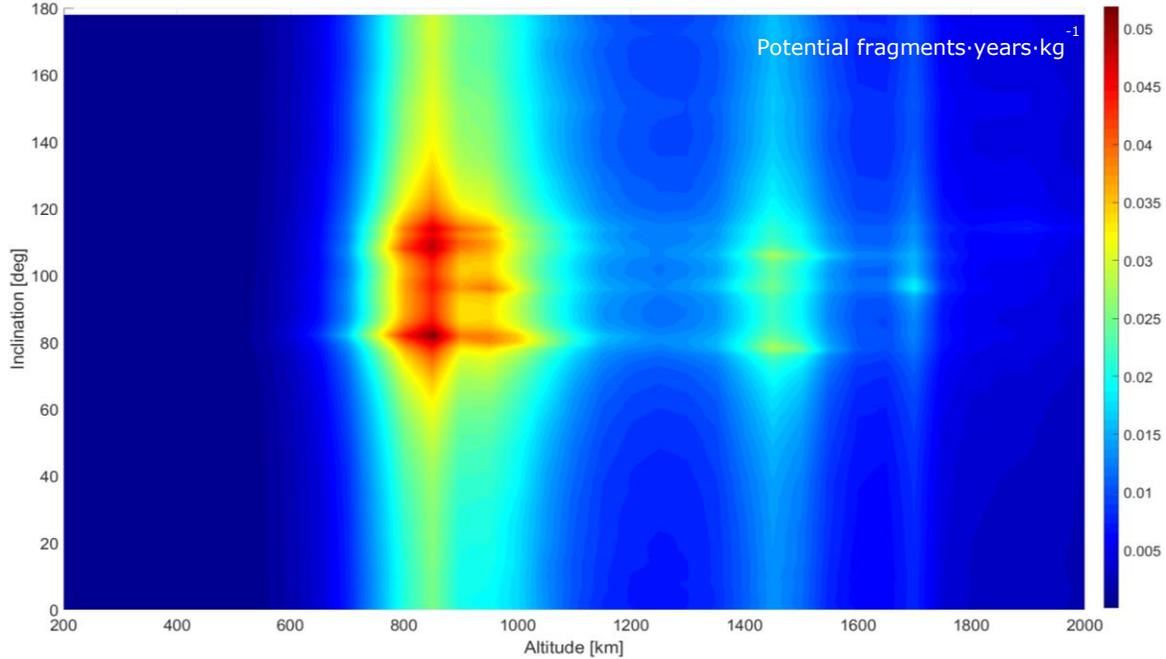

*Figure 6. Potential fragments-year per kg generated as characterisation factors for the LCA indicator. The value of $\alpha$ equals 0,86 fragments·kg⁻¹ as the interval [1.000-1.500 kg] is considered for the spacecraft. Calculations are made for each of the 3330 discretised circular orbits.*

## 3 RESULTS

### 3.1 Baseline Sentinel-3 environmental profile

The relative contribution of the different stages under study are presented in Figure 7. The detailed definition, qualification and production stage contributes by around 50% for most of the impact categories, except *mineral resource depletion potential* (~100%) and *space debris impact category* (0%). Within the design and production phase, the office work (i.e. electricity and natural gas consumption) is by far the major contributor (>50% for all the impact categories) except for the *mineral resource depletion* driven by the materials used for satellites. This score (~100%) is lead by the production of the PV system (for solar cells) using Germanium which is the most important semiconductor material, in weight, of a spacecraft. Assembly, Integration and Testing (AIT) accounts for around 20% of the global warming potential within the production stage. The scores related to the *toxicity/ecotoxicity categories* are driven by the production of the electronic components embedded in the payload that account for around 30%.

The utilisation stage of the spacecraft contributes to 25 to 35%% of the environmental impacts for the overall categories including *space debris generation potential*. During this phase, the electricity consumption of the data centre and servers are majors contributors for the *global warming potential, toxicity and ecotoxicity potential as well as fossil resource depletion.*

Others life cycle steps (*i.e. phases A+B, E1a and F*) have a low contribution to the conventional LCA impact categories. Electricity consumption for office work (related to R&D) during pre-design and design stages has a rather low contribution to the environmental indicators, mainly because of the highly decarbonised electricity mix in France and for the ESA Technical centre (where 100% of renewable electricity is used). It is also the case for the *Phase E1a*, which includes launch and early orbit phase (LEOP) and commissioning with activities at ESA Operation Center (100% renewable electricity too).

Finally, the disposal phase contribution equals to zero on all conventional indicators because of the lack of model related to the emissions on the Earth surface and into the atmosphere during the spacecraft re-entry.

Regarding the space debris related impacts (i.e. relative score on degradative use of orbits), 36% of the impact comes from the *utilisation phase*. The potential annual impact during the operational phase is more important than the *disposal phase*. It is mainly due to the coordinates of the operational orbit (i.e. SSO, 800 km, 98°) for which the *exposure factor* faces one the highest value in the LEO region, even considering only 1cm<debris size<10cm to take into account collision avoidance manoeuvres for bigger debris. Also, the lack of efficient atmospheric drag at such altitude gives a high value in term of *severity factor*. Nevertheless, the small amount of propellant available at the EoL of the mission (17 kg) does not allow to target a 25-years reentry. As a consequence, the long duration of the disposal phase (around 89 years) compared to the initial lifetime of the mission results in a higher score (64%) of the disposal phase even if the *exposure & severity factors* of the orbits crossed during the PMD are lower.

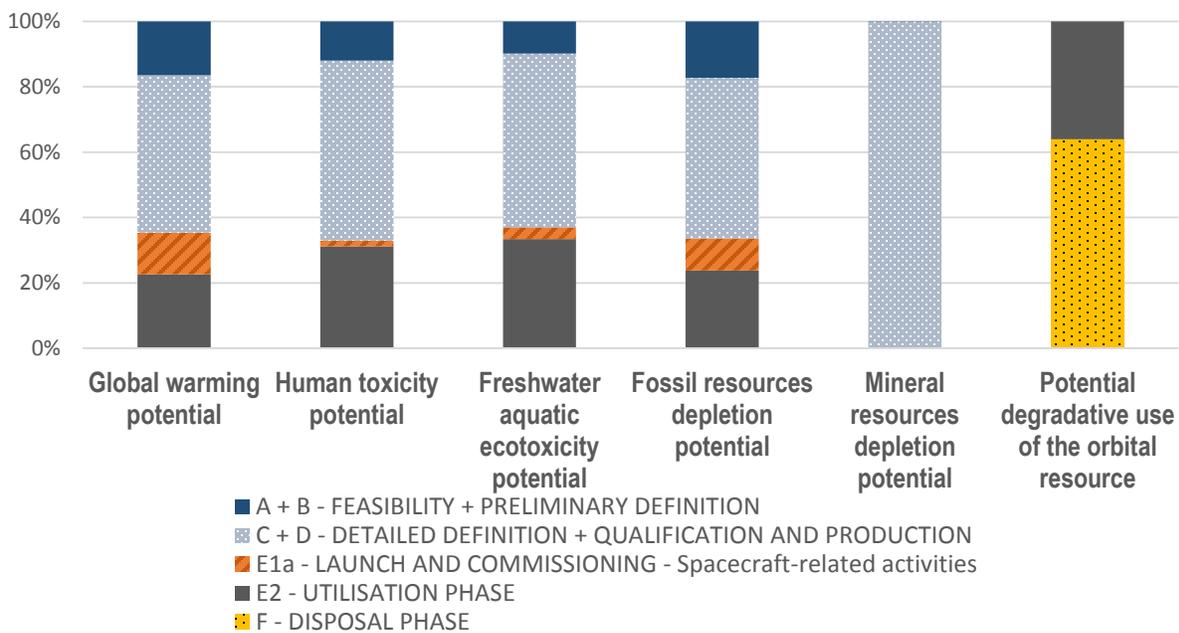

*Figure 7. Environmental profile of the Sentinel-3 mission, including the relative contributions of utilisation and disposal phases in term of space debris potential. Impact scores for conventional categories are retrieved from* [10].

### 3.2 Comparison of the impact score for the baseline and 'GreenSat' considering the 'degradative use' of the orbital resource

Figure 8 shows the potential number of *fragments·years* generated by both scenarios. We observe that the impact is 2.3 times higher for the baseline scenario which considers 120 kg of hydrazine than with the redesigned 'GreenSat' scenarios using 165 kg of LMP-103s. Alternatively, we also compute the probability of collision only considering the relevant XFs. We obtained the following probabilities of collision (conservative approach): *6,6%, 10,5% and 27,4%*, respectively for a 12,5-year mission with direct deorbiting (discarded option due to the delta-V budget), with a delayed reentry in less than 25 years (i.e. 'GreenSat') and with a lowering perigee manoeuvre with remaining propellant (i.e. Baseline scenario). Besides, we notice that, in the case of GreenSat, the contribution of the *utilisation phase* corresponds to around 87% of the overall impact (versus 36% previously). It is due to a shorter PMD lifetime in the orbital environment while the operational time remains the same.

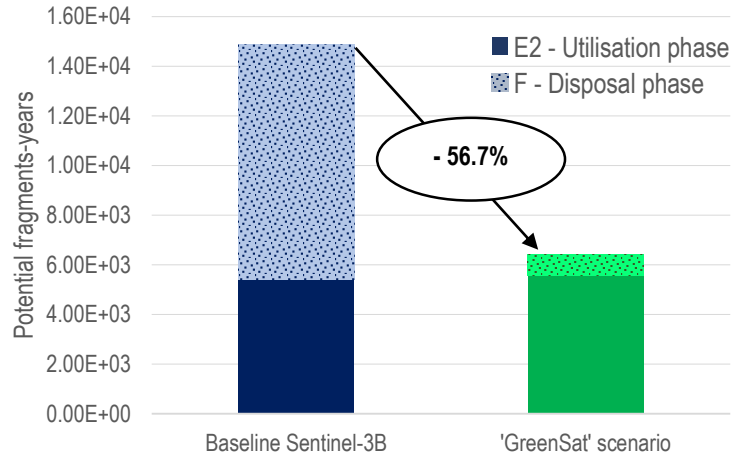

*Figure 8. Comparison of the impact scores obtained for the baseline (i.e. PMD duration 89 years) and the 'GreenSat' scenarios (i.e. PMD <25 years). The difference between both "E2 – Utilisation phase" comes from the launch mass increased in the 'GreenSat' scenario (1195 kg vs 1150 kg for the baseline scenario with hydrazine).*

### 3.3 Global comparison of the LCA results

While the replacement of hydrazine by LMP-103s significantly reduces the potential impact on the orbital environmental, the scores obtained for conventional environmental impacts in both scenarios are retrieved from [10] and compared in Figure 9. The results show that no significant *burden shifting* (i.e.transferring environmental impact from one category to another) occurs. Even if the impacts related manufacturing stage is higher in the case of the new LMP-103s propellant (due to the larger amount of propellant by +32%), the difference when considering the full scope of the study is compensated mainly by a shortened launch preparation into clean rooms. It stems from more limited safety measures regarding 'GreenSat' scenario during the loading propellant operations and cleaning processes. Indeed, hydrazine is targeted by the REACH regulation, which is not the case for the LMP-103s.

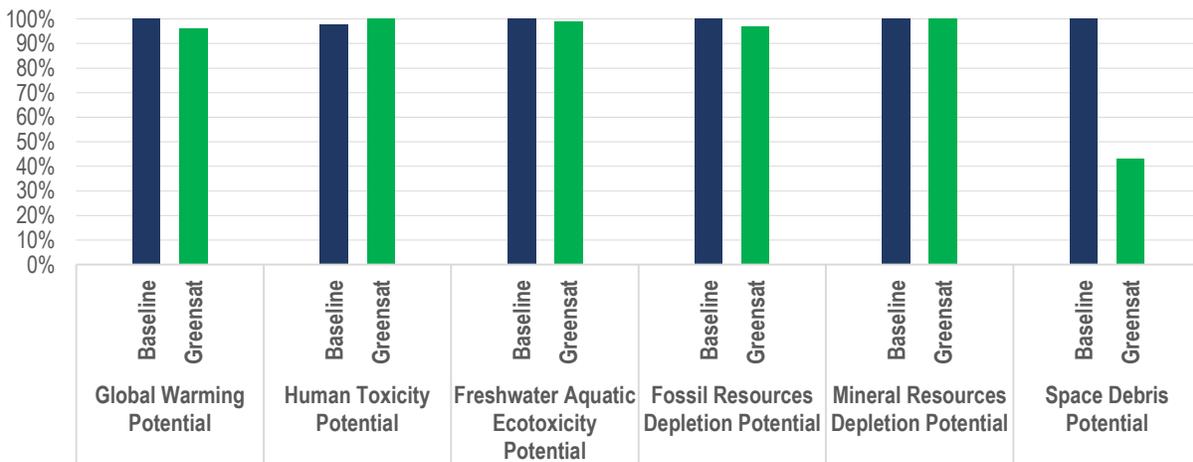

*Figure 9. Comparative environmental impacts of Sentinel-3B at mission level for Baseline and 'GreenSat scenario'. The figure highlights the absence of a noticeable 'burden shifting' between the impact categories. Hence, the GreenSat scenario can be considered as a relevant ecodesign improvement.*

## 4 CONCLUSION & PERSPECTIVES

This LCA study demonstrates the added value of holistically addressing the environmental performance of space mission, *i.e.* not only in term of conventional environmental impacts but also assessing the potential impact on the orbital environment. Based on the previous results obtained in the frame of the 'GreenSat' ecodesign study and the scores computed for the *potential degradative use* of the orbital environment, the replacement of the Hydrazine by LMP-103s propellant appears as a relevant ecodesign improvement for the sentinel-3 mission. The full environmental profile computed represents a relevant input to be used by decision-makers at the early design stage of space mission.

Having more propellant dedicated to the PMD would also help the satellite to initiate a controlled re-entry aiming to comply with the casualty risk threshold required by the space debris mitigation standards. Nevertheless, additional investigations should be carried out for a detailed comparison of both propulsion systems. Sentinel-3 spacecraft was designed with "light" 1N hydrazine thrusters, while LMP-103s propulsion system could require new thruster equipment and addition of a re-pressurisation module to cope with the needs of a controlled re-entry.

Going further, relevant mission parameters could be taken into account to perform a comprehensive multi-criteria optimisation when designing space missions and the associated spacecraft. For instance, minimum performance thresholds should be considered when defining the functional unit of the system. It is particularly the case for the reliability rate (e.g. 90% reliability at the end of the mission) as it could greatly influence the use phase and the EoL durations (e.g. when mission extensions are decided) [30]. It is also the case for the casualty risk related to the atmospheric re-entry as mentioned above.

Finally, as stated by Donella Meadows [31]: "*indicators arise from values (we measure what we care about), and they create values (we care about what we measure)*". The need of consistent metrics for space sustainability assessment is a raising concern among the aerospace community as demonstrated by the on-going work of Letizia et al. [32]. In this context, the LCA methodology should be considered as a starting point guiding ecodesign effort for space systems in a broader space sustainability perspective.

## 5 ACKNOWLEDGEMENTS

The authors acknowledge the support of the French National Association for Technical Research (CIFRE Convention 2015/1269). The research on the LCA indicator for space debris was also cofounded by the R&T department of the ArianeGroup in the frame of the Eco-space project.

We also acknowledge the project COMPASS "Control for Orbit Manoeuvring through Perturbations for Application to Space Systems" (Grant agreement No 679086) funded by the European Research Council (ERC) under the European Union Horizon 2020 research and innovation programme.

The authors are grateful of the work previously performed by Thales Alenia Space and Deloitte Consulting in the frame of the ESA funded project 'GreenSat' (ESA-TEC-SOW-003752).## 6 REFERENCES

[1] CEOS, ESA, Satellite earth observations in support of climate information challenges, 2015. http://eohandbook.com/cop21/files/CEOS_EOHB_2015_COP21.pdf.

[2] T. Maury, P. Loubet, S.M. Serrano, A. Gallice, T. Maury, P. Loubet, S.M. Serrano, A. Gallice, Application of environmental Life Cycle Assessment ( LCA ) within the space sector: a state of the art, Acta Astronaut. (2020). https://doi.org/10.1016/j.actaastro.2020.01.035.

[3] J. Guinée, R. Heijungs, G. Huppes, A. Zamagni, P. Masoni, R. Buonamici, T. Ekvall, T. Rydberg, Life cycle assessment: past, present, and future., Environ. Sci. Technol. 45 (2011) 90–96. https://doi.org/10.1021/es101316v.

[4] A. Gallice, T. Maury, E. Olmo, Environmental impact of the exploitation of the Ariane 6 launcher system, in: Clean Sp. Ind. Days, ESA - Clean Space, ESTEC, Noordwijk, 2018. https://indico.esa.int/event/234/contributions/3918/attachments/3115/3826/2018CSID_AGallice_EnvironmentalLifeCycleImpactAnalysisOfA6ExploitationPhase.pdf.

[5] T. Maury, P. Loubet, M. Trisolini, A. Gallice, G. Sonnemann, C. Colombo, Assessing the impact of space debris on orbital resource in life cycle assessment: A proposed method and case study, Sci. Total Environ. (2019). https://doi.org/10.1016/j.scitotenv.2019.02.438.